# Interfacial and Electronic Properties of Heterostructures of MXene and Graphene

Rui Li,[1,2] Weiwei Sun,[3,4] Cheng Zhan,[5] Paul R. C. Kent,[3,4] and De-en Jiang[2,*]

[1]Department of Chemistry, Liaocheng University, Liaocheng, Shandong 252059, China
[2]Department of Chemistry, University of California, Riverside, California 92521, US
[3]Computational Sciences and Engineering Division, Oak Ridge National Laboratory, Oak Ridge, Tennessee 37831, US
[4]Center for Nanophase Materials Sciences, Oak Ridge National Laboratory, Oak Ridge, Tennessee 37831, US
[5]Quantum Simulation Group, Lawrence Livermore National Laboratory, Livermore, California 94550, US

*Corresponding author. Email: djiang@ucr.edu

Dated: 2 Feb. 2019

ABSTRACT:

MXene-based heterostructures have received considerable interest owing to their unique properties. Herein, we examine various heterostructures of the prototypical MXene $Ti_3C_2T_2$ (T=O, OH, F; terminal groups) and graphene using density functional theory. We find that the adhesion energy, charge transfer, and band structure of these heterostructures are sensitive not only to the surface functional group, but also to the stacking order. Due to its greatest difference in work function with graphene, $Ti_3C_2(OH)_2$ has the strongest interaction with graphene, followed by $Ti_3C_2O_2$ and then $Ti_3C_2F_2$. Electron transfers from $Ti_3C_2(OH)_2$ to graphene but from graphene to $Ti_3C_2O_2$ and $Ti_3C_2F_2$, which causes a shift in the Dirac point of the graphene bands in the heterostructures of monolayer graphene and monolayer MXene. In the heterostructures of bilayer graphene and monolayer MXene, the interface breaks the symmetry of the bilayer graphene; in the case of the AB-stacking bilayer, the electron transfer leads to an interfacial electric field that opens up a gap in the graphene bands at the K point. This internal polarization strengthens both the interfacial adhesions and the cohesion between the two graphene layers. The MXene-graphene-MXene and graphene-MXene-graphene sandwich structures behave as two mirror-symmetric MXene-graphene interfaces. Our first principles studies provide a comprehensive understanding for the interaction between a typical MXene and graphene.

# I. INTRODUCTION

Heterostructures of different two-dimensional (2D) materials, such as graphene, hexagonal BN (h-BN), MoS$_2$, and phosphorene, are attracting greater attention [1-3]. These heterostructures, held together mainly by van der Waals (vdW) forces, can mix the intrinsic electronic properties of the dissimilar 2D materials and result in new electronic properties that may have potential applications in electric energy storage, electronics, and catalysis. MXenes are a new family of 2D transition-metal carbides/carbonitrides/nitrides/borides that have already shown promises in batteries, capacitors, and electrocatalysis owing to their diverse and attractive properties [4-12]. Hence, the large variety of compositions provide a new type of building block beyond the usual 2D materials (such as graphene, h-BN, and MoS$_2$) for composite materials [13-21].

The weak Fermi-level pinning at 2D vdW heterostructures of metal-semiconductor junctions has been shown to enable the effective tuning of Schottky barrier for 2D metals such as h-NbS$_2$ [22]. Since most functionalized MXenes are metallic [23, 24] and their work function can be tuned by surface termination [25-27], using MXenes as the 2D metal has the potential to yield novel designs of 2D vdW heterostructures of metal-semiconductor junctions with tunable Schottky barrier heights for electronics applications.

The composite materials of MXenes have gathered more interest in capacitive energy storage. As a representative MXene, Ti$_3$C$_2$T$_x$ (T = OH, O and F; terminal groups) has been experimentally investigated for its excellent performance in capacitive cycling [28-33]. Aïssa *et al.* [15] investigated the transport properties of a sandwich-like Ti$_3$C$_2$T$_x$/graphene composite and found a clear correlation of both electrical conductance and Hall carrier mobility with respect to the graphene concentration. Xu *et al.* [16] showed that Ti$_3$C$_2$T$_x$/graphene films exhibited a high volumetric capacitance and a synergistic effect between graphene and Ti$_3$C$_2$T$_x$ nanosheets. Yan *et al.* [17] demonstrated that graphene sheet could be inserted into the MXene layers to form a well-aligned ordered structure, preventing the self-restacking of MXene layers and facilitating the rapid diffusion and transport of electrolyte ions in the increased interlayer spacing.

Despite the many experimental studies on the Ti$_3$C$_2$T$_x$/graphene composites and layered structures, the fundamental interfacial energetic and electronic properties of the MXene/graphene heterostructures are still unclear. Here we investigate the $Ti_3C_2T_x$/graphene heterostructures by means of first-principles density functional theory (DFT). We consider different configurations of stacking, including $Ti_3C_2T_x$/graphene, graphene/$Ti_3C_2T_x$/graphene, $Ti_3C_2T_x$/graphene/$Ti_3C_2T_x$, and comparison of graphene single layer and bilayer for interfacing $Ti_3C_2T_x$. We focus on interfacial charge transfer and adhesion energetics as well as the change in the band structure with respect to the individual building blocks.

We organize the rest of the paper as follows: Section II describes the models for the $Ti_3C_2T_x$/graphene heterostructures and the DFT method we use; Sec. III presents the results of geometries, band structures, and energetics as well as their analyses; Sec. IV summarizes the main results and conclusions.

## II. HETEROSTRUCTURE MODELS AND COMPUTATIONAL DETAILS

We consider three different surface-termination groups for $Ti_3C_2T_2$, with T = O, OH, and F, namely, $Ti_3C_2O_2$, $Ti_3C_2(OH)_2$, and $Ti_3C_2F_2$; for convenience, they are abbreviated as MO, MOH, and MF, respectively. All the functional groups are located at the fcc hollow site of $Ti_3C_2$ which is energetically more favorable than the hcp site [24]. The calculated in-plane lattice constants are 3.033 Å, 3.081 Å, 3.076 Å, and 2.458 Å for MO, MOH, MF and graphene, respectively. To minimize the lattice mismatch, a (4×4) supercell of $Ti_3C_2T_2$ is matched to a (5×5) supercell of graphene, which yields small lattice mismatches of 1.3%, 0.1%, and 0.3% for MO, MOH, and MF on graphene, respectively. For the heterostructures of $Ti_3C_2T_2$ MXene (M) and graphene (G), five stacking patterns are examined: (i) monolayer MXene and monolayer graphene (M_G); (ii) monolayer MXene and AA-stacking bilayer graphene (M_2$G_{AA}$); (iii) monolayer MXene and AB-stacking bilayer graphene (M_2$G_{AB}$); (iv) graphene-MXene-graphene sandwich structure (G_M_G); (v) MXene-graphene-MXene sandwich structure (M_G_M).

First principles calculations are carried out using density functional theory (DFT) as implemented in the Vienna ab initio simulation package (VASP) [34] with periodic boundary conditions. The projector-augmented wave (PAW) method [35] and GGA-PBE [36] exchange–correlation functional are used. The Grimme's DFT-D3 scheme [37] of dispersion correction with zero damping is adopted to account for the van der Waals (vdW) interactions. The cutoff energy for the plane-wave basis set is 500 eV. A vacuum layer of 15 Å is applied to minimize the interaction of the heterostructure slab and its periodic images along the $z$ direction. The Brillouin zone is sampled with a (4×4×1) k-mesh within the Monkhorst–Pack scheme. The structures are fully optimized using convergence criteria of $10^{-5}$ eV for the total energy and 0.02 eVÅ$^{-1}$ for forces. The adhesive energy of the interface between M and G is defined as $E_{Inter} = (E_M + E_G - E_{total})/A$, where $E_{total}$, $E_M$ and $E_G$, represent the energies of the interfacial system, the MXene layer, and the graphene layer, respectively; A is the area of the interface. In the sandwich structures, there are two interfaces between MXene and graphene; the average adhesive energy of the two interfaces is evaluated as the following: $E_{Inter} = (2E_M + E_G - E_{total})/2A$ for M_G_M and $E_{Inter} = (E_M + 2E_G - E_{total})/2A$ for G_M_G.

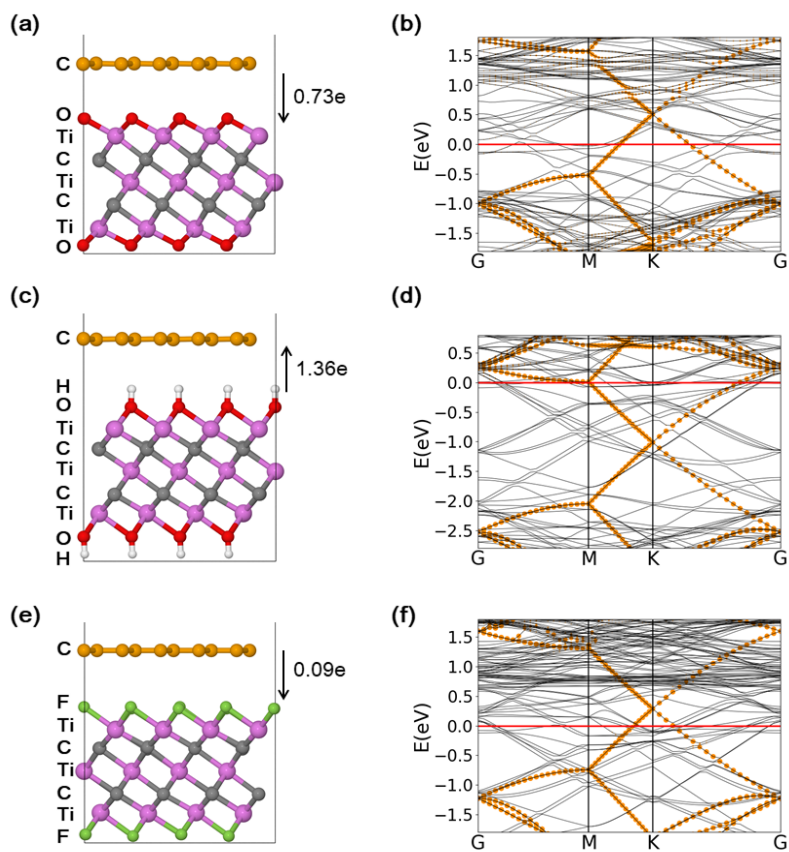

FIG. 1. Optimized geometries and band structures of the monolayer Ti$_3$C$_2$T$_2$_graphene heterostructure: (a) and (b), Ti$_3$C$_2$O$_2$_graphene; (c) and (d), Ti$_3$C$_2$(OH)$_2$_graphene; (e) and (f), Ti$_3$C$_2$F$_2$_graphene. The direction and number of electrons transferred between MXene and graphene is also given. The brown points in the band structures stand for the projected band structures of graphene and their sizes represent the magnitude of contribution. Fermi level is indicated by a red line in the band structure.

## III. RESULTS AND DISCUSSIONS

To understand the interfacial energetic and electronic properties, we investigate the Ti$_3$C$_2$T$_2$/graphene heterostructures in different stacking order and their interfacial charge transfer, adhesion energetics, and band structure. We start with the simple monolayer MXene-graphene heterostructures and compare their properties with the individual building blocks and then move to more complex interfaces.

### A. Monolayer MXene with monolayer graphene

Fig. 1 shows the optimized geometries and band structures of the monolayer heterostructures of Ti$_3$C$_2$O$_2$_graphene, Ti$_3$C$_2$(OH)$_2$_graphene, and Ti$_3$C$_2$F$_2$_graphene. For convenience, we abbreviate them as MO_G, MOH_G, and MF_G, respectively. In MOH_G, the interlayer distance is 2.8 Å, smaller than that of MO_G (3.1 Å) or MF_G (3.2 Å), suggesting a stronger interlayer interaction in MOH_G. The computed interfacial adhesive energies are shown in Table I. Indeed, MOH_G has the strongest interlayer adhesion (3.60 eV/nm$^2$), followed by MO_G (2.75 eV/nm$^2$) and MF_G (1.97 eV/nm$^2$). A previous report has shown that typical adhesion energies in 2D vdW heterostructures are round 20 meV/ Å$^2$ or 2.0 eV/nm$^2$ [38]. Indeed, we found that the adhesion energies at the interfaces between MF and G as well as between two MXene layers (Table I) are close to this value, indicating the dominance of the vdW interaction. In contrast, the much higher adhesion energies found for the MO_G and MOH_G indicate additional contribution due to the interfacial charge transfer.

From the computed Bader charges, we found that the interlayer interaction correlates with the amount of charge transfer across the interface (Fig. 1): 1.36 electron from MOH to G, 0.73 electron from G to MO, and 0.09 electron from G to MF. The different direction of electron transfer in MOH_G from in MO_G and MF_G prompted us to compute the work functions of the individual building blocks. As shown in Table II, the work function of the MXene is very sensitive to the surface termination groups. While the work functions of MO and MF are higher than that of graphene, MOH's is much lower. The work-function difference therefore dictates the direction and degree of electron transfer across the interface, after proper alignment of the vacuum levels in the heterostructure. Electrons consequently transfer from graphene to MXene in MO_G and MF_G, but from MXene to graphene in MOH_G; and the number of transferred electrons correlates with the difference in work function. The ultra-low work function of MOH has been attributed to the surface dipole of the O-H groups [25, 39]. Modulation of the work function by surface engineering has been previously suggested for MXene design [26].

TABLE I. Interfacial adhesive energy ($E_{Inter}$) between MXene and graphene in all heterostructures examined in this work. MO=$Ti_3C_2O_2$; MOH=$Ti_3C_2(OH)_2$; MF= $Ti_3C_2F_2$; G=graphene; 2G=graphene-bilayer.

| Structure | $E_{Inter}$ (eV/nm$^2$) | Structure | $E_{Inter}$ (eV/nm$^2$) | Structure | $E_{Inter}$ (eV/nm$^2$) |
|---|---|---|---|---|---|
| MF_G | 1.97 | MF_2G$_{AA}$ | 2.17 | G_MF_G | 1.99 |
| MO_G | 2.75 | MO_2G$_{AA}$ | 3.04 | G_MO_G | 2.72 |
| MOH_G | 3.60 | MOH_2G$_{AA}$ | 3.88 | G_MOH_G | 3.55 |
| MF_MF | 1.75 | MF_2G$_{AB}$ | 2.15 | MF_G_MF | 2.17 |
| MO_MO | 2.01 | MO_2G$_{AB}$ | 3.09 | MO_G_MO | 2.90 |
| MOH_MOH | 1.89 | MOH_2G$_{AB}$ | 3.98 | MOH_G_MOH | 3.52 |

TABLE II. The calculated work function (eV) of the building blocks of the heterostructures

| Building block | Work function (eV) |
|---|---|
| $Ti_3C_2O_2$ | 5.97 |
| $Ti_3C_2(OH)_2$ | 2.02 |
| $Ti_3C_2F_2$ | 4.78 |
| Monolayer graphene | 4.22 |
| Bilayer graphene AA stacking | 4.26 |
| Bilayer graphene AB stacking | 4.24 |

The interfacial charge transfer is also reflected in the band structure (Fig. 1). Despite the band mixing in all three heterostructures, the main feature of graphene is preserved but shifted. In MO_G and MF_G, the bands and the Dirac point of graphene are shifted upward by 0.52 eV and 0.25 eV, respectively, relative to the Fermi level, because of MO and MF's higher work functions and the electron transfer from graphene to MXene, while in MOH_G the bands and the Dirac point of graphene are shifted downward by 0.98 eV because of MOH's lower work function and the electron transfer from MXene to graphene. Fig. 2 shows the electron-density plot of the heterostructure minus the sum of the individual layers. One can see that the charge transfer happens predominately at the interface: in MO_G [Fig. 2(a)], the π electrons from graphene are mainly moved to O atoms on MO, while in MOH_G [Fig. 2(b)], the electrons from -OH groups on MOH are shifted to the π orbitals of graphene and located mainly at the interface below the graphene sheet. In MF_G [Fig. 2(c)], the extent of transfer is much less.

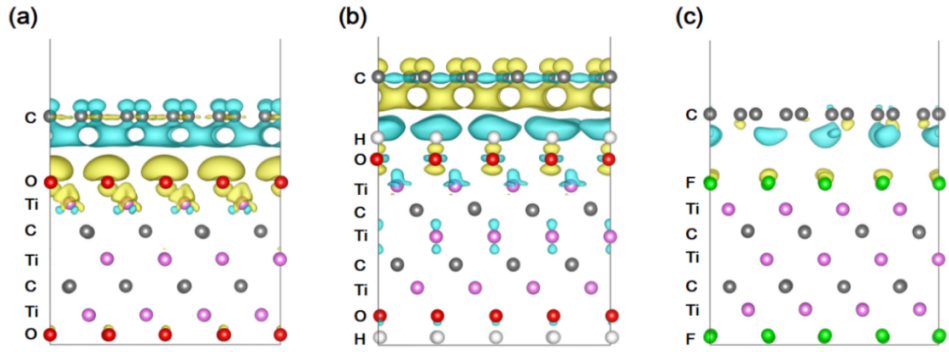

FIG. 2. Electron-density difference at the interface of (a) $Ti_3C_2O_2$_graphene (MO_G), (b) $Ti_3C_2(OH)_2$_graphene (MOH_G), and (c) $Ti_3C_2F_2$_graphene (MF_G). The excess (yellow) and depleted (cyan) electrons are shown: the isosurface value is $4\times10^{-4}$ e/Å$^3$ for (a) and (b); $1.5\times10^{-4}$ e/Å$^3$ for (c).

**B. Monolayer MXene with bilayer graphene**

Bilayer graphene allows us to explore how the MXene/graphene interface impacts the two layers of graphene and their band structures differently with the monolayer of graphene. We considered both AB-stacking and AA-stacking for the bilayer graphene and abbreviate them as $2G_{AB}$ and $2G_{AA}$. For example, a heterostructure of the $Ti_3C_2O_2$ MXene with the AA-stacking graphene bilayer will be denoted as MO_$2G_{AA}$. The optimized geometries and band structures of the MXene_2G heterostructures are shown in Fig. 3 and Fig. 4 for AA and AB stacking, respectively. Since the work function of the bilayer graphene is similar to that of the monolayer graphene (Table II), the direction and extent of electron transfer across the interface are similar for heterostructures of MXene_2G and MXene_G. The difference lies in how the total amount of electron transfer is distributed between the two graphene layers, e.g., homogenously or inhomogeneously. Our finding [Figs. 3(a), 3(d) and 3(g)] is that the graphene layer closer to MXene makes significantly more contributions than the farther layer, due to its stronger interaction with the MXene layer.

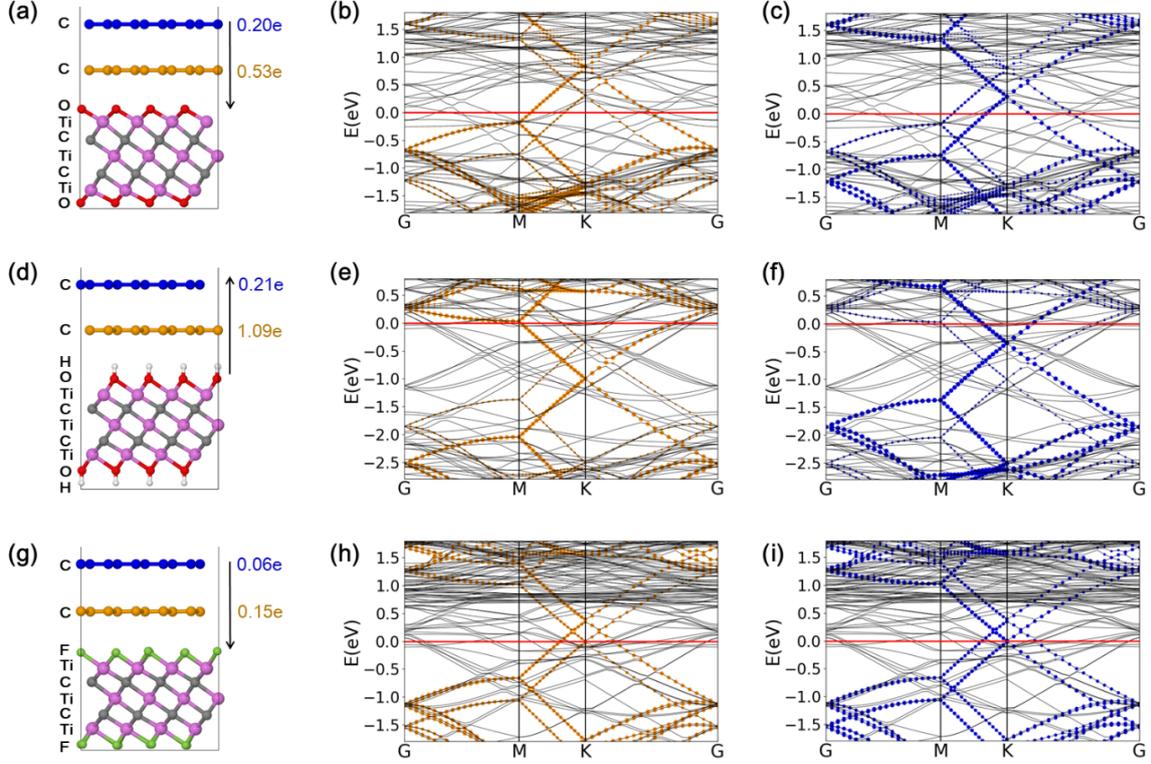

FIG. 3. Optimized geometries and band structures of the heterostructures of $Ti_3C_2T_2$ with AA-stacking bilayer graphene ($2G_{AA}$): (a)-(c), $Ti_3C_2O_2\_2G_{AA}$; (d)-(f), $Ti_3C_2(OH)_2\_2G_{AA}$; (e)-(i), $Ti_3C_2F_2\_2G_{AA}$. The direction and number of electron transfer between MXene and graphene bilayer are also given. The brown (blue) points in the band structures stand for the projected band structures of the lower (upper) graphene layer; their sizes represent the magnitude of contribution. The Fermi level is indicated by a red line in the band structure.

The symmetry breaking of the graphene bilayer also manifests in the band structures. In $Ti_3C_2T_2\_2G_{AA}$ heterostructures, the main features of the energy bands of $2G_{AA}$ are preserved [40], especially for $Ti_3C_2F_2\_2G_{AA}$ [Figs. 3(h) and 3(i)] where the interfacial interaction is weak. There are two sets of the subbands because the degeneracy of two graphene layers. The linear bands are observed symmetrically around the K point intersection. But due to the MXene layer, the two subbands are shifted with different degrees. For example, in $MO\_2G_{AA}$, the set of subbands and its Dirac point from mainly the closer graphene layer are shifted up further [Fig.

3(b)] than those from the farther layer [Fig. 3(c)] above the Fermi level. In MOH_2G$_{AA}$ [Figs. 3(e) and 3(f)], the bands are shifted below the Fermi level.

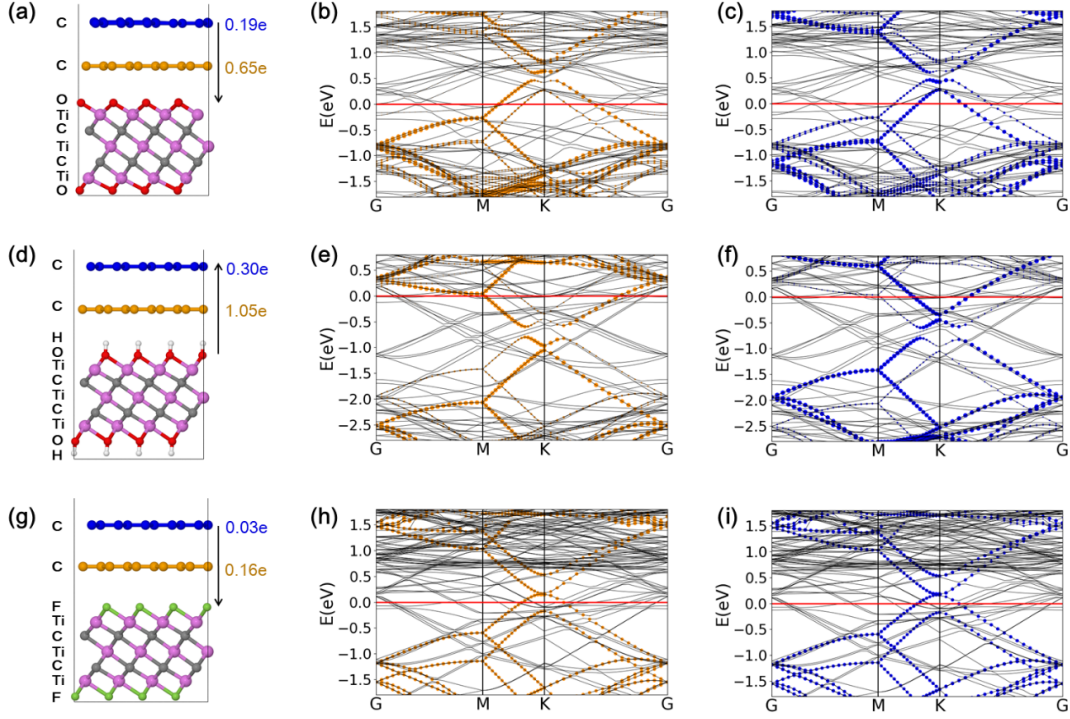

FIG. 4. Optimized geometries and band structures of the heterostructures of Ti$_3$C$_2$T$_2$ with AB-stacking bilayer graphene: (a)-(c), Ti$_3$C$_2$O$_2$_2G$_{AB}$; (d)-(f), Ti$_3$C$_2$(OH)$_2$_2G$_{AB}$; (e)-(i), Ti$_3$C$_2$F$_2$_2G$_{AB}$. The direction and number of electron transfer between MXene and graphene bilayer are also given. The brown (blue) points in the band structures stand for the projected band structures of the lower (upper) graphene layer; their sizes represent the magnitude of contribution. The Fermi level is indicated by a red line in the band structure.

The Ti$_3$C$_2$T$_2$_2G$_{AB}$ heterostructures show very different band structures from Ti$_3$C$_2$F$_2$_2G$_{AA}$ for the graphene part. The difference in the band structure between AA and AB stackings of pristine bilayer graphene has been previously studied by others [41-44], and our results are consistent with theirs. Briefly, the interlayer coupling and the asymmetry in the AB stacking destroy the symmetry of the bands and causes the linear bands to become parabolic near the K-point. Due to the high symmetry in the AA stacking, the interlayer coupling just shifts the linear bands horizontally and oppositely off the K point.

For Ti$_3$C$_2$T$_2$_2G$_{AB}$ heterostructures, the main band features of pristine 2G$_{AB}$ can still be found in MF_2G$_{AB}$ [Figs. 4(h) and 4(i)]: two nearly parallel parabolic π* bands locate above two nearly parallel parabolic π bands with nearly zero band gap at the K point [41, 43]; the electronic distribution of the two layers of graphene is uniform in the two sets of subbands which are slightly shifted upward ~0.2 eV from the Fermi level. However, in MO_2G$_{AB}$ [Figs. 4(b) and 4(c)] and MOH_2G$_{AB}$ [Figs. 4(e) and 4(f)]: a sizable gap develops at the K point between the upper and lower bands which now show a Mexican-hat shape due to the interaction of the two layers. Such features were also found in 2G$_{AB}$ in an external electric field [43, 45]. Similarly, one can consider that 2G$_{AB}$ in MO_2G$_{AB}$ and MOH_2G$_{AB}$ is under an internal polarization after interfacial electron transfer, leading to charge imbalance in the two graphene layers and the band-structure features.

TABLE III. The cohesive energy (E$_{G-G}$) between two layers of graphene in MXene_2G heterostructures in comparison with that of pristine graphene bilayer (2G) for both AA and AB stacking configurations.

| Structure | E$_{G-G}$ (eV/nm$^2$) | Structure | E$_{G-G}$ (eV/nm$^2$) |
|---|---|---|---|
| 2G$_{AA}$ | 1.40 | 2G$_{AB}$ | 1.59 |
| MF_2G$_{AA}$ | 1.69 | MF_2G$_{AB}$ | 1.80 |
| MO_2G$_{AA}$ | 1.81 | MO_2G$_{AB}$ | 1.98 |
| MOH_2G$_{AA}$ | 1.84 | MOH_2G$_{AB}$ | 2.00 |

TABLE IV. The distance (D$_{G-G}$) between two layers of graphene in MXene_2G heterostructures in comparison with that in pristine graphene bilayer (2G) for both AA and AB stacking configurations.

| Structure | D$_{G-G}$ (Å) | Structure | D$_{G-G}$ (Å) |
|---|---|---|---|
| 2G$_{AA}$ | 3.53 | 2G$_{AB}$ | 3.33 |
| MF_2G$_{AA}$ | 3.54 | MF_2G$_{AB}$ | 3.35 |
| MO_2G$_{AA}$ | 3.53 | MO_2G$_{AB}$ | 3.33 |
| MOH_2G$_{AA}$ | 3.53 | MOH_2G$_{AB}$ | 3.34 |

In comparison with the monolayer MXene_G heterostructures, the interfacial adhesion is about 10% stronger in the MXene_2G heterostructures (Table I). Comparing AA and AB stackings, we found that the strength of the interfacial adhesion is similar for MF_2G but slightly stronger for AB stacking in MO_2G and MOH_2G. Moreover, the cohesive energy between the two graphene layers ($E_{G-G}$) is in fact larger in MXene_2G heterostructures than in pristine 2G (Table III) for both AA and AB stackings. We think that this enhanced cohesion is due to the polarization of the graphene bilayer by the MXene layer in the heterostructures; the interlayer spacing between the two graphene layers remains similar (Table IV), indicating similar strength of van der Waals attraction.

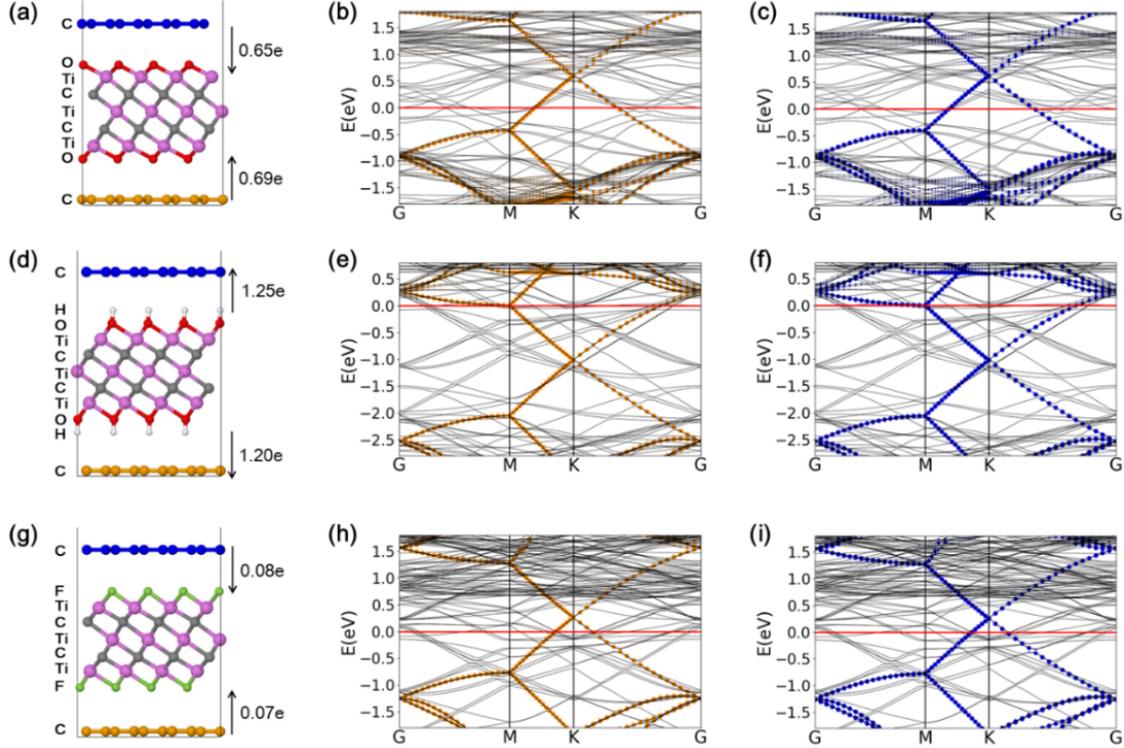

FIG. 5. Optimized geometries and band structures of the graphene-$Ti_3C_2T_2$-graphene sandwich structures. (a)-(c), G_$Ti_3C_2O_2$_G; (d)-(f), G_$Ti_3C_2(OH)_2$_G; (e)-(i), G_$Ti_3C_2F_2$_G. Direction and number of electron transfer between MXene and graphene layers are also given. The brown (blue) points in the band structures stand for the projected band structures of the lower (upper) graphene layer; their sizes represent the magnitude of contribution. Fermi level is indicated by a red line in the band structure.

Comparing the cohesions of graphene layers (1.59 eV/nm$^2$ for AB-stacking graphene bilayer; Table III) and the MXene bilayers (1.75 eV/nm$^2$ for MF_MF; 2.01 eV/nm$^2$ for MO_MO; 1.89 eV/nm$^2$ for MOH_MOH; Table I) with the interfacial adhesions of the heterostructures (Table I), one can see that forming the M_G interface is in fact thermodynamically more favorable for all three MF_G, MO_G, and MOH_G interfaces. It is therefore expected that graphene can be intercalated between MXene layers and vice versa, leading to, for example, the sandwich structures which we examine next.

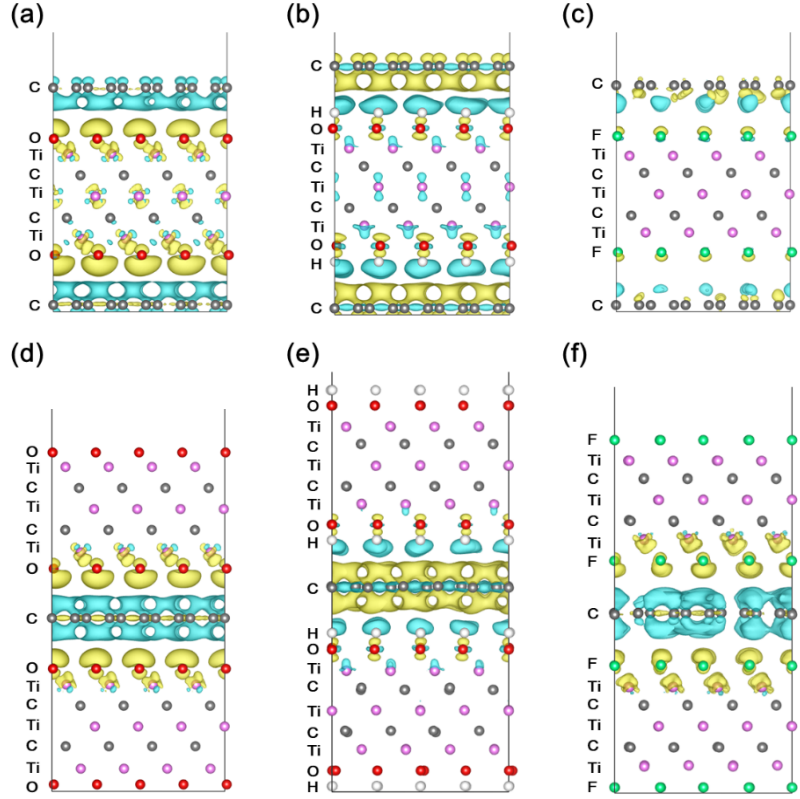

FIG. 6. Electron-density difference at the interface of (a) G_MO_G, (b) G_MOH_G, (c) G_MF_G (d) MO_G_MO, (e) MOH_G_MOH, and (f) MF_G_MF. The excess (yellow) and depleted (cyan) electrons are shown: the isosurface value is $4\times10^{-4}$ e/Å$^3$ for (a) and (b); $1.5\times10^{-4}$ e/Å$^3$ for (c).

**C. Graphene_MXene_graphene sandwich structures**

Sandwich structures allow us to explore two MXene-graphene interfaces at the same time. Fig. 5 shows the optimized geometries and band structures of the graphene-$Ti_3C_2T_2$-graphene heterostructures for the three terminal groups, denoted as G_MO_G, G_MOH_G, and G_MF_G. The electron transfer occurs in both the upper and lower interfaces; the two interfaces behave very similarly, as seen in both the amount of electron transfer and the band structure. Essentially, the graphene-$Ti_3C_2T_2$-graphene heterostructure can be viewed as two separate $Ti_3C_2T_2$-graphene interfaces, so the interfacial adhesion energy and the band structure (Table I and Fig. 5) of graphene-$Ti_3C_2T_2$-graphene are similar to those of $Ti_3C_2T_2$-graphene (Table I and Fig. 1), even though the amount of single-interface electron transfer is slightly smaller in graphene-Ti$_3$C$_2$T$_2$-graphene. This is because the MXene layer needs to accept or donate electrons to two graphene layers. Indeed, the electron-density-difference plots [Figs. 6(a) and 6(b)] show that the Ti atoms in the middle of the MXene layer are also involved slightly in the electron transfer.

**D. MXene-graphene-MXene sandwich structures**

In contrast with the graphene-MXene-graphene sandwich structures where the two MXene-graphene interfaces are relatively separated and can be essentially viewed as two individual MXene-graphene interfaces, the graphene monolayer in MXene-graphene-MXene is shared by two MXene layers. The two types of sandwich structures share some common features, such as the mirror-symmetric nature of charge transfer and band structures against the middle plane (Fig. 7). In other words, the MXene-graphene-MXene sandwich structure can still be viewed as two separate MXene-graphene interfaces. The difference is that the graphene layer now provides or accepts more electrons, leading to greater up [Fig. 7(b)] or down [Fig. 7(d)] shifts of the graphene bands and the Dirac point. Compared with the graphene-MXene-graphene sandwich structures, the average MXene-graphene adhesion in MXene-graphene-MXene structures is stronger for MO and MF.

It is worth noting that there is a much larger electron transfer between MF and graphene in MF_G_MF (~0.16) than in MF_G (0.09) or G_MF_G (~0.07). We think that this is due to the polarization effect of MF to the graphene layer's electrons both below and above its atomic plane in MF_G_MF. For MF_G [Fig. 2(c)] and G_MF_G [Fig. 6(c)], the electron depletion is mainly concentrated below the graphene plane facing MF while there is little change to the electrons above the graphene layer facing the vacuum. In contrast, the electrons both below and above the graphene plane are activated under the influence of polarization in MF_G_MF [Fig. 6(f)]; in other words, the synergy and close proximity of the two interfaces above and below the graphene plane reinforce each other in facilitating the electron transfer. Such synergy is obviously missing in MF_G and G_MF_G.

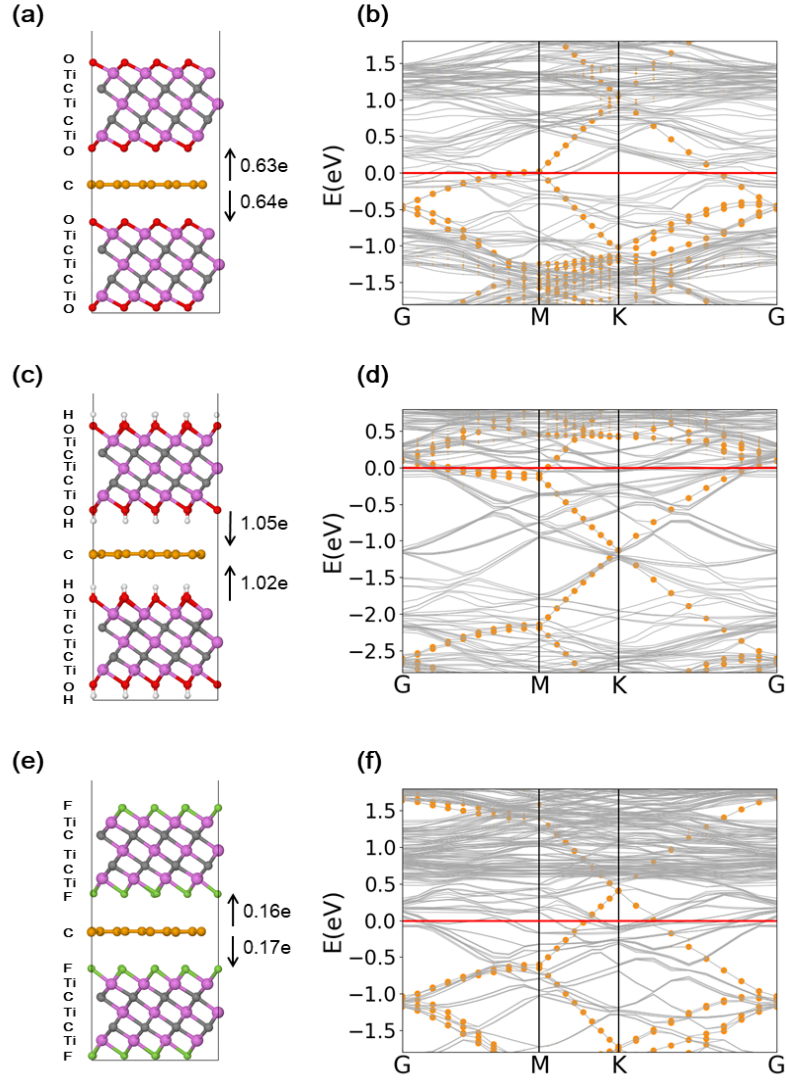

FIG. 7. Optimized geometries and band structures of the $Ti_3C_2T_2$_graphene_$Ti_3C_2T_2$ sandwich structures. (a)-(b), $Ti_3C_2O_2$_G_$Ti_3C_2O_2$; (c)-(d), $Ti_3C_2(OH)_2$_G_$Ti_3C_2(OH)_2$; (e)-(f), $Ti_3C_2F_2$_G_$Ti_3C_2F_2$. The direction and number of electron transfer between MXene and graphene layers are also given. The brown points in the band structures stand for the projected band structures of the graphene layer; their sizes represent the magnitude of contribution. Fermi level is indicated by a red line in the band structure.

## IV. CONCLUSION

In sum, we have carried out a comprehensive first principles study of the structural and electronic properties of $Ti_3C_2T_2$/graphene heterostructures of different configurations for T=O, OH, and F groups. We found that the difference in work functions of the individual building layers dictates the direction and magnitude of electron transfer at the interface. The resulting polarization of the interface impacted both the strength of the interfacial adhesion and the band structure. It also broke the symmetry of the bilayer graphene in the heterostructures, but strengthened both the interfacial adhesions and the graphene cohesion. The MXene-graphene-MXene and graphene-MXene-graphene sandwich structures could be understood as two mirror-symmetric separate MXene-graphene interfaces.

## ACKNOWLEDGMENTS

This research is sponsored by the Fluid Interface Reactions, Structures, and Transport (FIRST) Center, an Energy Frontier Research Center funded by the U.S. Department of Energy (DOE), Office of Science, Office of Basic Energy Sciences. R.L. was supported by the Natural Science Foundation of Shandong Province (ZR2017QB010).

## REFERENCES

[1] A. K. Geim, and I. V. Grigorieva, Nature **499**, 419 (2013).
[2] K. S. Novoselov, A. Mishchenko, A. Carvalho, and A. H. Castro Neto, Science **353**, aac9439 (2016).
[3] E. Pomerantseva, and Y. Gogotsi, Nat. Energy **2**, 17089 (2017).
[4] M. Naguib, O. Mashtalir, J. Carle, V. Presser, J. Lu, L. Hultman, Y. Gogotsi, and M. W. Barsoum, ACS Nano **6**, 1322 (2012).
[5] M. Naguib, V. N. Mochalin, M. W. Barsoum, and Y. Gogotsi, Adv. Mater. **26**, 992 (2014).
[6] B. Anasori, Y. Xie, M. Beidaghi, J. Lu, B. C. Hosler, L. Hultman, P. R. C. Kent, Y. Gogotsi, and M. W. Barsoum, ACS Nano **9**, 9507 (2015).
[7] B. Anasori, M. R. Lukatskaya, and Y. Gogotsi, Nat. Rev. Mater. **2**, 16098 (2017).
[8] N. K. Chaudhari, H. Jin, B. Kim, D. San Baek, S. H. Joo, and K. Lee, J. Mater. Chem. A **5**, 24564 (2017).
[9] H. Tang, Q. Hu, M. Zheng, Y. Chi, X. Qin, H. Pang, and Q. Xu, Prog. Nat. Sci. **28**, 133 (2018).
[10] H. Zhang, H. Xiang, F.-z. Dai, Z. Zhang, and Y. Zhou, J. Mater. Sci. Technol. **34**, 2022 (2018).
[11] Z. Jiang, P. Wang, X. Jiang, and J. Zhao, Nanoscale Horiz. **3**, 335 (2018).
[12] Z. Guo, J. Zhou, and Z. Sun, J. Mater. Chem. A **5**, 23530 (2017).
[13] L.-Y. Gan, Y.-J. Zhao, D. Huang, and U. Schwingenschlögl, Phys. Rev. B **87**, 245307 (2013).
[14] Y. Gao, L. Wang, Z. Li, A. Zhou, Q. Hu, and X. Cao, Solid. State Sci. **35**, 62 (2014).
[15] B. Aïssa, A. Ali, K. A. Mahmoud, T. Haddad, and M. Nedil, Appl. Phys. Lett. **109**, 043109 (2016).
[16] S. Xu, G. Wei, J. Li, W. Han, and Y. Gogotsi, J. Mater. Chem. A **5**, 17442 (2017).
[17] J. Yan, C. E. Ren, K. Maleski, C. B. Hatter, B. Anasori, P. Urbankowski, A. Sarycheva, and Y. Gogotsi, Adv. Funct. Mater. **27**, 1701264 (2017).
[18] Q. Yang, Z. Xu, B. Fang, T. Huang, S. Cai, H. Chen, Y. Liu, K. Gopalsamy, W. Gao, and C. Gao, J. Mater. Chem. A **5**, 22113 (2017).
[19] S. Cao, B. Shen, T. Tong, J. Fu, and J. Yu, Adv. Funct. Mater. **28**, 1800136 (2018).
[20] J. Low, L. Zhang, T. Tong, B. Shen, and J. Yu, J. Catal. **361**, 255 (2018).
[21] S. Zhou, X. Yang, W. Pei, N. Liu, and J. Zhao, Nanoscale **10**, 10876 (2018).

[22] Y. Liu, P. Stradins, and S.-H. Wei, Science Advances **2**, e1600069 (2016).
[23] M. Khazaei, M. Arai, T. Sasaki, C.-Y. Chung, N. S. Venkataramanan, M. Estili, Y. Sakka, and Y. Kawazoe, Adv. Funct. Mater. **23**, 2185 (2013).
[24] C. Zhan, W. Sun, P. R. C. Kent, M. Naguib, Y. Gogotsi, and D. E. Jiang, J. Phys. Chem. C **123**, 315 (2019).
[25] M. Khazaei, M. Arai, T. Sasaki, A. Ranjbar, Y. Liang, and S. Yunoki, Phys. Rev. B **92**, 075411 (2015).
[26] H. A. Tahini, X. Tan, and S. C. Smith, Nanoscale **9**, 7016 (2017).
[27] Y. Liu, H. Xiao, and W. A. Goddard, J. Am. Chem. Soc. **138**, 15853 (2016).
[28] L. Yu, L. Hu, B. Anasori, Y.-T. Liu, Q. Zhu, P. Zhang, Y. Gogotsi, and B. Xu, ACS Energy Lett. **3**, 1597 (2018).
[29] X. Xie, K. Kretschmer, B. Anasori, B. Sun, G. Wang, and Y. Gogotsi, ACS Appl. Nano Mater. **1**, 505 (2018).
[30] C. J. Zhang, B. Anasori, A. Seral-Ascaso, S. H. Park, N. McEvoy, A. Shmeliov, G. S. Duesberg, J. N. Coleman, Y. Gogotsi, and V. Nicolosi, Adv. Mater. **29**, 1702678 (2017).
[31] M. R. Lukatskaya, S. Kota, Z. Lin, M.-Q. Zhao, N. Shpigel, M. D. Levi, J. Halim, P.-L. Taberna, M. W. Barsoum, P. Simon, and Y. Gogotsi, Nat. Energy **2**, 17105 (2017).
[32] M. Boota, B. Anasori, C. Voigt, M. Q. Zhao, M. W. Barsoum, and Y. Gogotsi, Adv. Mater. **28**, 1517 (2016).
[33] C. Zhan, M. Naguib, M. Lukatskaya, P. R. C. Kent, Y. Gogotsi, and D. E. Jiang, J. Phys. Chem. Lett. **9**, 1223 (2018).
[34] G. Kresse, and J. Furthmüller, Phys. Rev. B **54**, 11169 (1996).
[35] G. Kresse, and D. Joubert, Phys. Rev. B **59**, 1758 (1999).
[36] J. P. Perdew, K. Burke, and M. Ernzerhof, Phys. Rev. Lett. **77**, 3865 (1996).
[37] S. Grimme, J. Antony, S. Ehrlich, and H. Krieg, J. Chem. Phys. **132**, 154104 (2010).
[38] T. Bjorkman, A. Gulans, A. V. Krasheninnikov, and R. M. Nieminen, Phys. Rev. Lett. **108**, 235502 (2012).
[39] Y. Liu, H. Xiao, and W. A. Goddard III, J. Am. Chem. Soc. **138**, 15853 (2016).
[40] I. Lobato, and B. Partoens, Phys. Rev. B **83**, 165429 (2011).
[41] Y. Nam, D.-K. Ki, D. Soler-Delgado, and A. F. Morpurgo, Science **362**, 324 (2018).
[42] A. V. Rozhkov, A. O. Sboychakov, A. L. Rakhmanov, and F. Nori, Phys. Rep. **648**, 1 (2016).
[43] T. Wang, Q. Guo, Y. Liu, and K. Sheng, Chin. Phys. B **21**, 067301 (2012).
[44] Y. Xu, X. Li, and J. Dong, Nanotechnology **21**, 065711 (2010).
[45] G. M. Rutter, S. Jung, N. N. Klimov, D. B. Newell, N. B. Zhitenev, and J. A. Stroscio, Nat. Phys. **7**, 649 (2011).